\documentclass[aps,prb,twocolumn,superscriptaddress,showpacs,amsmath,amssymb,longbibliography]{revtex4-2}
\usepackage{graphicx}
\usepackage[colorlinks=true,citecolor=blue,linkcolor=blue,urlcolor=blue,bookmarks=true]{hyperref}

\usepackage[utf8]{inputenc}
\usepackage[english]{babel}
\usepackage{amsmath}
\usepackage{xcolor}
\usepackage{soul}

\definecolor{nred} {RGB}{224,0,0}
\definecolor{nblue} {RGB}{28,130,185}
\definecolor{dgreen}{RGB}{78,138,21}
\definecolor{norange}{RGB}{230,120,20}

\begin{document}

\title{Finding local integrals of motion in quantum lattice models \\ in the thermodynamic limit}  
\author{J. Paw\l{o}wski}
\affiliation{Institute of Theoretical Physics, Faculty of Fundamental Problems of Technology, Wroc\l{a}w University of Science and Technology, 50-370 Wroc\l{a}w, Poland}
\author{J. Herbrych}
\affiliation{Institute of Theoretical Physics, Faculty of Fundamental Problems of Technology, Wroc\l{a}w University of Science and Technology, 50-370 Wroc\l{a}w, Poland}
\author{M. Mierzejewski}
\affiliation{Institute of Theoretical Physics, Faculty of Fundamental Problems of Technology, Wroc\l{a}w University of Science and Technology, 50-370 Wroc\l{a}w, Poland}

\begin{abstract} 
Local integrals of motion (LIOMs) play a key role in understanding the long-time properties of closed macroscopic systems. They were found for selected integrable systems via complex analytical calculations. The existence of LIOMs and their structure can also be studied via numerical methods, which, however, involve exact diagonalization of Hamiltonians, posing a bottleneck for such studies. We show that finding LIOMs in translationally invariant lattice models or unitary quantum circuits can be reduced to a problem for which one may numerically find an exact solution in the thermodynamic limit.  We develop a simple algorithm and demonstrate its efficiency by calculating LIOMs and bounds on correlations (the Mazur bounds) for infinite integrable spin chains and unitary circuits. Finally, we demonstrate that this approach identifies slow modes in nearly integrable spin models and estimates their relaxation times.
\end{abstract}

\maketitle

\section{Introduction}

Closed macroscopic quantum systems undergo a unitary evolution. However, a reduced density matrix of a subsystem is expected to reach a stationary state. The latter is determined by local conservation laws or, equivalently, by local integrals of motion (LIOMs). In particular, the expectation values of local observables are equal to averages over an appropriate ensemble~\cite{DAlessio2016}.
 Generic  systems have only a  few  conserved quantities (like Hamiltonian,  particle number, magnetization) and {the relevant ensemble has the form of} the thermal Gibbs state \cite{Goldstein2006, Linden2009,polkovnikov2011}.  Interacting integrable systems have extensive number of LIOMs and the stationary states {(of subsystems)} have the form that is consistent with the generalized Gibbs ensemble~\cite{Rigol2007,Kollar2011,Cassidy2011,Gogolin2011,Pozsgay2013,Fagotti2014,Mierzejewski2014,Pozsgay2014,Goldstein2014,Ilievski2015,Langen2015,Vidmar2016,Palmai2018,Fukai2020,Reiter2021,Vernier2023}. Identification of LIOMs is also the starting point for construction of the hydrodynamics or the generalized hydrodynamics~\cite{Bertini2016a,Castro-Alvaredo2016,Ilievski2017,Doyon2017,Gopalakrishnan2018, DeNardis2018,Agrawal2020,Friedman2020,Bastianello2021,Gopalakrishnan2023a,Hubner2025a,Doyon2025} in  case of integrable models. Moreover, anomalous transport (or its anticipated absence) in various quantum systems has recently been linked to the presence of LIOMs.  Most prominent examples concern strongly disordered systems~\cite{Serbyn2013,Huse2014,Chandran2015,ros2015,Rademaker2016,Inglis2016,Monthus2016,Imbrie2017a,Mierzejewski2018,Goihl2018,Kulshreshtha2018,Krajewski2022a},  tilted chains with emergent conservation of the dipole moment~\cite{Lydzba2024} and systems exhibiting the Hilbert space fragmentation~\cite{Moudgalya2020,Bastianello2022,Lydzba2024}.
   
While the importance of LIOMs for the long-time dynamics appears unquestionable,   in many cases  their construction remains a challenging problem. Even for the most studied integrable interacting model, i.e., the XXZ chain,  the complete set of LIOMs (that includes also quasilocal LIOMs) has been established analytically only recently~\cite{Prosen2011a,Prosen2013a,Prosen2014a, Pereira2014,Ilievski2015,Ilievski2015a,Ilievski2016}. In many cases, LIOMs are identified through complex calculations that are specific to the investigated model~\cite{tetelman1982a,Shastry1986,Olmedilla1988,Grabowski1995,Yang2022,Fukai2023a,  Yamaguchi2024,Shiraishi2025}. Although there are model-independent approaches which allow to find LIOMs in integrable models~\cite{Mierzejewski2015a,Ulcakar2024} or  approximate LIOMs in nearly integrable systems~\cite{Mierzejewski2015}, they are based on exact diagonalization of Hamiltonians. The latter poses serious limitations on the studied models and raises questions concerning the reliability of the finite-size scaling.     
  
In this work, we show that finding LIOMs in translationally-invariant lattice models belongs to the same class of problems as the high-temperature expansion \cite{Baker1967,Oitmaa1996,Laurent2024}. Namely, one can numerically find exact LIOMs and the corresponding Mazur bounds~\cite{mazur1969,zotos1997} in the thermodynamic limit without diagonalizing the Hamiltonian. This holds true for quantum systems where the dynamics is set either by a Hamiltonian or by a unitary transformation, as is the case in unitary circuits. The geometry of the system is not important, and the only essential assumption concerns the dynamics that needs to be set by local transformations.

\section{General approach}
\subsection{Algorithm}
We consider a system containing $L$ sites, which are enumerated by the index $l=1,\ldots,L$, and study extensive Hermitian operators, \mbox{$\hat O^s=\sum_l \hat O^s_l$}, where the upper index, $s$,  distinguishes between various operators. The densities,  $\hat O^s_l$, are local and act on up to $M$ sites adjacent to $l$. We introduce the Hilbert-Schmidt (HS) product $\langle \hat O^s \hat O^p \rangle ={\rm Tr}( \hat O^{s \dagger} \hat O^p)/(ZL)$ and the HS norm of operators, $\lVert\hat O^{s} \lVert^2=\langle \hat O^s \hat O^s \rangle$, where $Z$ is the dimension of the Hilbert space.  We recall that the HS product of the local densities, $\langle  \hat{O}^{s}_l  \hat{O}^{s'}_{l'} \rangle$, can be exactly calculated for infinite lattice systems using just the properties of the operator algebra, similarly to the technique employed in the high-temperature expansion.

Our main assumption is that the dynamics is set either by a local Hamiltonian, $\hat H$, or by a repeated sequence of local unitary transformations $\hat U$.  In the former case, we take $\hat H=\sum_l \hat H_l$, where  $\hat H_l$ acts on $M_H$ sites. We look for local operators, $\hat Q$, which commute with the Hamiltonian, $[\hat Q, \hat H]=0$ or, equivalently, we require vanishing of the norm of a Hermitian operator $\hat Q'=i [\hat Q, \hat H] $, i.e., $\lVert\hat Q'\lVert^2=\langle \hat Q' \hat Q' \rangle=0$. We note that $\hat Q'$ is still local, $\hat Q'=\sum_l  \hat Q'_l$, where $\hat Q'_l$ acts on up to $M+M_H-1$ sites. For this reason, $\langle \hat Q' \hat Q' \rangle $ can also be calculated exactly in the thermodynamic limit. In case of circuits built out of repeated sequence of local unitary transformations, the same reasoning can be applied for \mbox{$\hat Q'=\hat U^{\dagger}\hat Q \hat U-\hat Q$}.  I.e., $\hat Q'$ is a local operator and vanishing norm of $\hat Q'$ implies that $\hat Q$ is conserved. Both cases can be comprehended within a unified formulation. Namely, we introduce a function $f(\hat O^s)$ which has the following properties: {\it (i)} $f$ is linear, {\it (ii)} vanishing of $\lVert f(\hat O^s)\lVert$ is equivalent to conservation of $\hat O^s$; {\it (iii)} $f(\hat O^s)$ is a local and Hermitian operator so that one can exactly calculate $\langle f(\hat O^s)f(\hat O^p)\rangle $. If the dynamics is set by a Hamiltonian then one may choose $f(\hat  O^s)=i[\hat O^s,\hat H]$, whereas $f(\hat O^s)=\hat U^{\dagger}\hat O^s \hat U-\hat O^s$ can be used for unitary circuits. The specific choice of $f$ is irrelevant and can be adapted to the desired setup, as long as it is consistent with conditions {\it (i)}-{\it (iii)}.

As a next step, we consider a set containing $N_O$ orthonormal operators, $\langle \hat O^s \hat O^{p} \rangle=\delta_{sp}$. We aim to find {\em all} LIOMs which can be expressed as linear combinations of these operators. To this end, we solve the eigenproblem
\begin{equation}
\sum_{sp} V_{\alpha s} F_{sp} \; (V^T)_{p\beta}=
\lambda_{\alpha} \delta_{\alpha \beta},
\label{eig}
\end{equation} 
where $V$ is an orthogonal matrix and $F$ is a matrix of HS products, $F_{sp}=\langle f(\hat O^s) f(\hat O^p) \rangle $, constructed in the basis (of size $N_O$) containing all $\hat O^s$.
Since $f$ is a linear function, one may introduce rotated operators
\begin{equation}  
\hat A^{\alpha}=\sum_s V_{\alpha s} \hat O^s, \label{rot}
\end{equation}
and rewrite \eqref{eig} as
\begin{equation}
 \langle f(\hat A^{\alpha}) f(\hat A^{\beta}) \rangle =\lambda_{\alpha} \delta_{\alpha \beta}, \label{eig1} 
\end{equation}
from which one finds the norm $\lVert f(\hat A^{\alpha})\lVert=\sqrt{\lambda_{\alpha}}$.
The rotated operators are local and orthogonal, $\langle  \hat A^{\alpha} \hat A^{\beta} \rangle=\delta_{\alpha \beta}$. Consequently, all $\hat A^{\alpha}$ corresponding to the eigenvalue $\lambda_{\alpha}=0$ represent orthogonal LIOMs, identified here via the norm $\lVert f(\hat A^{\alpha})\lVert=0$.  To distinguish these LIOMs from other operators introduced in \eqref{rot}, we denote LIOMs as $\hat Q^{\alpha}$ where $\alpha=1,\dots,N_L$ and $N_L$ is the number of LIOMs given by the degeneracy of the eigenvalue $\lambda_{\alpha}=0$ in \eqref{eig}. We keep the notation $\hat A^{\alpha}$,  for the remaining operators, $\alpha=N_L+1,...,N_O$ with
$\lambda_{\alpha}>0$.

As a final step, we show that the solution of \eqref{eig} identifies {\em all} LIOMs that can be represented via linear combinations of $\hat{O}^s$. To this end, we consider an operator $\hat C=\sum_{s} c_s \hat O^s$, $\lVert\hat{C}\lVert>0$, that is orthogonal to all $\hat Q^{\alpha}$ obtained from \eqref{eig}. We show  $\lVert f(\hat C)\lVert > 0$, implying that $\hat C$ is not conserved. Using the inverse of the  transformation from \eqref{rot}, such operator can be expressed in the rotated basis
\begin{eqnarray}
\hat C&=& \sum_{\alpha=1}^{N_L} \left(\sum_{s} c_s V_{\alpha s} \right) \hat Q^{\alpha}+\sum_{\alpha=N_L+1}^{N_O} \left(\sum_{s} c_s V_{\alpha s} \right) \hat A^{\alpha} \nonumber \\
&=&  \sum_{\alpha=1}^{N_L} \tilde{c}_{\alpha} \hat Q^{\alpha}+\sum_{\alpha=N_L+1}^{N_O} \tilde{c}_{\alpha} \hat A^{\alpha}. \label{beq}
\end{eqnarray}
The first term in \eqref{beq} vanishes since we assume that $\hat C$ is orthogonal to all $\hat Q^{\alpha}$. Then, one finds the squared norm
\begin{eqnarray}
\lVert f(\hat C) \lVert^2&=&\sum_{\alpha, \beta=N_L+1}^{N_O} \tilde{c}_{\alpha} \tilde{c}_{\beta} \langle f(\hat A^{\alpha}) f(\hat A^{\beta}) \rangle \nonumber \\ 
&=& \sum_{\alpha=N_L+1}^{N_O} \tilde{c}^{2}_{\alpha} \lambda_{\alpha} >0, \label{all}
\end{eqnarray}
where we have used \eqref{eig1}. The inequality in \eqref{all} follows from that all $\lambda_{\alpha}$ are positive for $\alpha > N_L$ and some $\tilde{c}_{\alpha}$ are nonzero (otherwise $\lVert\hat C\lVert=0$, contradicting one of the assumptions). 

{In what follows we demonstrate that  the eigenproblem in \eqref{eig} gives not only the LIOMs but also the Mazur bounds for {\em all}  operators $\hat O^s$ used in the construction of the $F$ matrix. To this end we start from a standard form of the Mazur bound~\cite{zotos1997}
\begin{equation}
 \lim_{T\to \infty} \frac{1}{T} \int_0^T {\rm d} t \,\langle \hat O^s(t) \hat O^s \rangle \ge \sum_{\alpha=1}^{N_L} \frac{\langle \hat O^s \hat Q^{\alpha} \rangle^2}{\langle \hat Q^{\alpha} \hat Q^{\alpha} \rangle},
 \label{mazur}
\end{equation}
and note that all \(\hat Q^{\alpha}\) are orthonormal, \(\langle \hat Q^{\alpha'} \hat Q^{\alpha} \rangle = \delta_{\alpha,\alpha'}\), so are
the operators $\hat O^s$, \mbox{$\langle \hat O^s \hat O^{s'} \rangle=\delta_{s,s'}$}.
Then, using Eq.~\eqref{rot} one may evaluate the expression on the right-hand side of Eq.~\eqref{mazur} 
\begin{equation}
    \sum_{\alpha=1}^{N_L} \frac{\langle \hat O^s \hat Q^{\alpha} \rangle^2}{\langle \hat Q^{\alpha} \hat Q^{\alpha} \rangle}= \sum_{\alpha=1}^{N_L} \sum_{s'} V^2_{\alpha s'} \langle \hat O^s \hat O^{s'} \rangle^2  = \sum_{\alpha=1}^{N_L} V^2_{\alpha s}.\
\end{equation}
}


For systems where the dynamics is set by the Hamiltonian, our approach is consistent with the idea from Ref. \cite{Huse2015,Bulchandani2022,Song2025} based on minimizing the norm of commutators $[\hat H, \hat A]$. For quantum circuits, it is similar (but not equivalent) to the approach from Refs. \cite{Prosen2002,Znidaric2024,znidaric2024a,Znidaric2025} based on Ruelle-Pollicott resonances obtained from diagonalization of a non-Hermitian matrix $\langle f(\hat O^s) \hat O^p \rangle$.

\subsection{The basis of operators}  
The first step in our algorithm is to choose the set of operators, $\hat O_s$, for which we then construct the matrix in \eqref{eig}. 
Any specific knowledge concerning LIOMs may help reduce the set of operators that are used for its construction. Without such knowledge, one may implement a brute-force approach and systematically include all operators with growing support.
In such case, one may use at each site, $l$, the set containing four operators: the identity matrix, $\hat \sigma^{0}_l$, and the Pauli matrices $\hat \sigma^{x}_l$, $\hat \sigma^{y}_l$, $\hat \sigma^{z}_l$. These operators are orthogonal, $\langle \hat \sigma^s_l \hat \sigma^{s'}_l\rangle=\delta_{s,s'}/L$, so they may be used for constructing extensive operators, $\hat O^{ s}=
\sum_l \hat O^{ s}_l$, supported on $m$ sites
\begin{equation}
\hat O^{ s}_l=\hat O^{\left(s_1,...,s_m\right)}_l=
\hat \sigma^{s_1}_{l} \hat \sigma^{s_2}_{l+1} ... \hat\sigma^{s_m}_{l+m-1}.
\label{dens}
\end{equation}
In the following, we assume that the first, $\hat \sigma^{s_1}_{l}$, and the last, $\hat \sigma^{s_m}_{l+m-1}$, spin operators are other than $\hat \sigma^{0}$. All operators with maximal support $M$ (i.e., generated for all $m\le M$) are orthogonal $\langle \hat O^{ s}_l \hat O^{ s'}_{l'}  \rangle=\delta_{l,l'} \delta_{ s, s'}/L$. {For fixed $M$, one can generate an exhaustive list by considering \(\hat O^s_l\) with the first operator chosen from the restricted set \(\{\hat \sigma^{x}_l, \hat \sigma^{y}_l, \hat \sigma^{z}_l\}\), and subsequent operators from the full set, including the identity. This guarantees no duplicates are present in the basis set after all the $N_O=3\times 4^{M-1}$ extensive operators are built. They are orthonormal $\langle \hat O^{ s} \hat O^{ s'} \rangle= \delta_{ s, s'}$ by construction}. We note that all translationally invariant operators supported on up to $M$ sites can be expressed as linear combinations of $\hat O^{s}$, thus one may easily calculate the HS product of such operators. Finally, we recall that product of two $\hat \sigma^s_l$-operators is proportional to other such operator, $\hat \sigma^s_l \hat \sigma^{s'}_l=\alpha_{s''} \hat \sigma^{s''}_l$, where 
$\alpha_{s''} \in \{1,\pm i\} $. This enables a straightforward calculation of $f(\hat O^{ s})$, which we implement using the software package from  Ref.~\cite{Sels2025}.  

\section{Results}
\subsection{LIOMs in the XXZ chain} 

\begin{figure}[hbtp]
    \centering
    \includegraphics[width=1.0\linewidth]{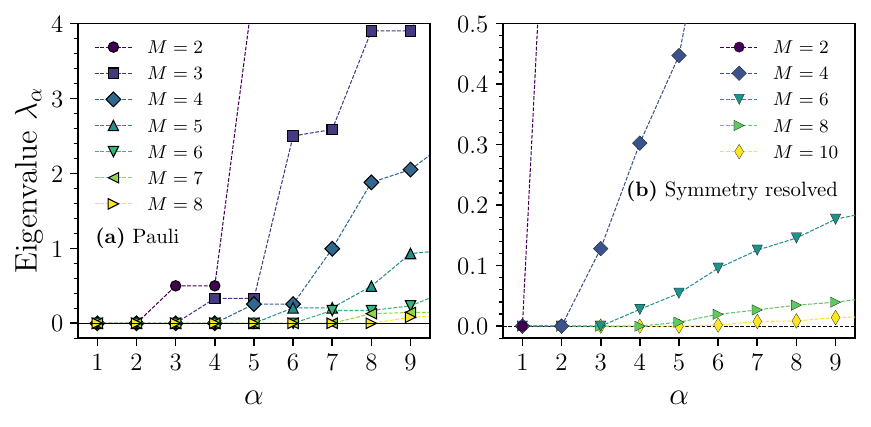}
    \caption{Lowest eigenvalues obtained from \eqref{eig} for the XXZ model with $\Delta=3/4$ and different operator bases: \textbf{(a)} Pauli string basis and \textbf{(b)} symmetry-resolved basis.}  
    \label{fig:xxz_lioms}
\end{figure}

The integrable XXZ chain appears as an obvious system for testing our method. The brute-force approach, as described above, allows to account for all operators supported on up to $M=8$ sites yielding $N_O=3 \times 4^7\sim 5 \times 10^4$ basis operators $\hat O^s$. To go beyond this limitation, we resolve the symmetries of the Hamiltonian (time-reversal, parity, conservation of~$\hat S^z_{\rm tot}=\sum_l \hat \sigma^z_l$), and note that  LIOMs can be found independently in each symmetry sector. Then,  it is convenient to use 
 \(\{\hat{\sigma}^0_l, \hat{\sigma}^+_l, \hat{\sigma}^-_l, \hat{\sigma}^z_l\}\) on-site basis, where  $ \hat{\sigma}^{\pm}_l=( \hat{\sigma}^x_l\pm i \hat{\sigma}^y_l)$. This basis is orthogonal but non-Hermitian
 so we first build the operators as in \eqref{dens} and then introduce the Hermitian counterparts,  
\(\hat O^{ {s}}_{l,R}=\hat O^{ {s}}_l + \hat O^{ {s} \dagger }_l\) and \(\hat O^{ {s}}_{l,I}=i(\hat O^{ {s}}_l - \hat O^{ {s} \dagger }_l)\).  We numerically solve the eigenproblem in \eqref{eig} for  $f[\hat  O^s_{R(I)} ]=i[\hat O^s_{R(I)},\hat H_{\text{XXZ}}]$ with the Hamiltonian
\begin{equation}
   \hat H_{\text{XXZ}} = \sum_{l=1}^L \frac{1}{2} \left[\hat \sigma_l^{+} \hat \sigma_{l+1}^{-} + \hat \sigma_l^{-} \hat \sigma_{l+1}^{+} \right] + \Delta \hat \sigma_l^z \hat 
   \sigma_{l+1}^z. 
   \label{hamxxz}
\end{equation} 
Fig.~\ref{fig:xxz_lioms}(a) shows the lowest $\lambda_{\alpha}$ obtained for $\Delta=3/4$ in the basis built from Pauli matrices (the Pauli string basis).  One observes that the eigenvalue $\lambda_{\alpha}=0$ is $M$-fold degenerate.  Figure~\ref{fig:xxz_lioms}(b) shows analogous results obtained in the symmetry-resolved basis, $\{ \hat O^{ {s}}_{R} \}$, containing operators that conserve total $\hat S^z$ and are even under time-reversal as well as the spin-flip (parity) symmetries. It reveals $M/2$ LIOMs in this symmetry sector, and we reach support up to $M=10$.  

 \begin{figure}[htbp]
    \centering
    \includegraphics[width=1\linewidth]{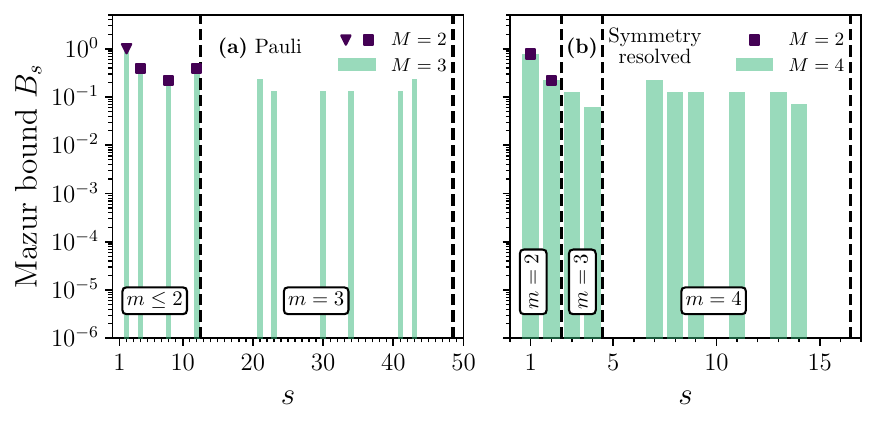}
    \caption{(a) and (b) shows the Mazur bounds \(B_s\) corresponding to LIOMs identified in Fig.~\ref{fig:xxz_lioms}(a) and \ref{fig:xxz_lioms}(b), respectively. 
   Index $s$ enumerates local
operators, $\hat O_s$, sorted accordingly to their support $m$.}
    \label{fig:mazur}
\end{figure}
 
While the degeneracy of the eigenvalue $\lambda_{\alpha}=0$  determines the number of LIOMs, the explicit form of LIOMs is encoded in the eigenvectors $V_{\alpha  s}$, see \eqref{rot}.  Eigenvectors within a degenerate subspace can be arbitrarily rotated so that the matrix $V_{\alpha  s}$ is not unique. One may choose a rotation, e.g., that restores the form of LIOMs known from the literature.  However, the Mazur bounds from \eqref{mazur}, $B_s=\sum_{\alpha=1}^{N_L} V^2_{\alpha s}$, are independent of such rotations. Therefore, the Mazur bounds are not only important for the dynamics of physically relevant operators, but they also show the structure of the LIOMs in a convenient and unique way. 

Figures~\ref{fig:mazur}(a) and \ref{fig:mazur}(b) show results for $B_s$ corresponding to LIOMs shown in Figs.~\ref{fig:xxz_lioms}(a) and \ref{fig:xxz_lioms}(b), respectively. The corresponding eigenvectors $\hat Q^{\alpha}$ represent the well-known integrals of motion of the XXZ chain.  In case of the Pauli string basis (left panel), triangle marks  $ \hat O^2=\hat S^z_{\rm tot}$ and squares mark the Hamiltonian containing $\hat O^4=
\sum_l \hat \sigma^x_l \hat \sigma^x_{l+1}$,
$\hat O^8=
\sum_l \hat \sigma^z_l \hat \sigma^z_{l+1}$, 
and 
$\hat O^{12}=
\sum_l \hat \sigma^y_l \hat \sigma^y_{l+1}$.
Both LIOMs are obtained for $M=2$. For larger support, $M=3$, one restores all LIOMs obtained for $M=2$ and, additionally, one also generates the energy current ($\hat Q^3$) marked as bars without symbols. In case of the symmetry resolved basis (right panel), one gets the Hamiltonian for $M=2$ (squares). For $M=4$ one obtains the Hamiltonian as well as $\hat Q^4$ (bars without symbols).  In Ref.~\cite{InfiniteLIOMs}, we provide simple codes, based on Ref.~\cite{Sels2025}, that generate LIOMs in the Pauli string basis as well as in the symmetry-resolved basis.

\subsection{Quasilocal conserved operators and spin stiffness}  
Here, we demonstrate that our approach also allows one to solve more demanding problem concerning quasilocal integrals of motion (QLIOMs) and their contribution to the Mazur bounds.  In contrast to LIOMs, QLIOMs have infinitely long tails in real space. Restricting QLIOMs to any finite support, $M$,
one obtains a quantity that is not strictly conserved, i.e., one obtains $\lambda_{\alpha}>0$. Despite being nonlocal, QLIOMs still do contribute to the Mazur bounds derived for local operator $\hat O^s$. One may formally write down both properties of QLIOMs using 
 Eqs.~(\ref{eig}) and (\ref{mazur}). Namely, $A^{\alpha}$ defined in \eqref{rot} is a QLIOM when 
\begin{eqnarray}
\lim_{M\to \infty} \lambda_{\alpha}&=& 0\; , \label{s_con} \\
\lim_{M\to \infty} V^2_{\alpha,s} &\ne& 0, \;\;\; {\rm for \; some \;} s.  \label{s_ql}
\end{eqnarray}
 We consider the XXZ chain, \eqref{hamxxz}, and study operators which are odd under the spin-flip transformation and odd under the time-reversal symmetry, $\hat O^{ s}_{I}=i(O^{  s}-O^{ s \dagger})$.  We have chosen this symmetry sector because it does not contain strictly local integrals of motion. Therefore, all contributions to the Mazur bounds must originate from QLIOMs. The simplest and also most studied observable in this symmetry sector is the spin current, $\hat J=\hat O^1=\sqrt{2}/4 \sum_l (i \hat \sigma^+_l \sigma^-_{l+1}+{\rm H.c.}) $. We introduce the prefactor, $\sqrt{2}/4$,  for which the spin current is normalized $\langle \hat J  \hat J \rangle=1$.

\begin{figure}[htbp]
    \centering
    \includegraphics[width=1.0\linewidth]{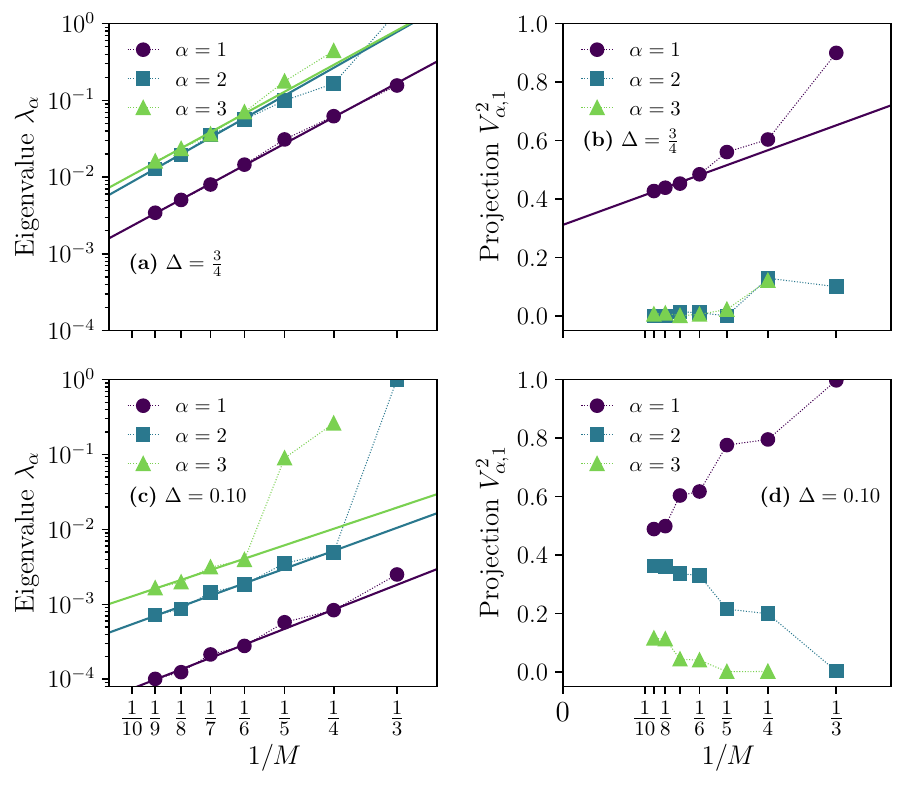}
    \caption{Three smallest eigenvalues, $\lambda_{\alpha}$, and eigenvectors, $V_{\alpha,1}$, versus maximal support, $M$, for XXZ model within symmetry resolved basis with (a),(b): $\Delta=3/4$
    and (c),(d),$\Delta=0.1$.} 
    \label{fig:quasilocal}
\end{figure}

Figure~\ref{fig:quasilocal}(a) shows three smallest eigenvalues obtained from solution of \eqref{eig} plotted vs. $1/M$ for \mbox{$\Delta=3/4$}. All three solutions, $\hat A^1,\hat A^2$ and $\hat A^3$, satisfy \eqref{s_con} thus they are conserved for $M \to \infty$. Figure~\ref{fig:quasilocal}(b) shows $M$-dependence of the corresponding eigenvectors, $V^2_{\alpha,1}$.  Based on these projections we  conclude that  $\hat A^1$  is quasilocal since $V^2_{1,1}$ satisfies \eqref{s_ql}. Quasilocality of $\hat A^2$ and $\hat A^3$, i.e., the  \eqref{s_ql}, can be further tested for other operators $\hat O^{s}\ne \hat O^1$. Figures~\ref{fig:quasilocal}(c) and \ref{fig:quasilocal}(d) show similar results for small anisotropy, $\Delta=0.1$. 

\begin{figure}[htbp]
    \centering
    \includegraphics[width=1.0\linewidth]{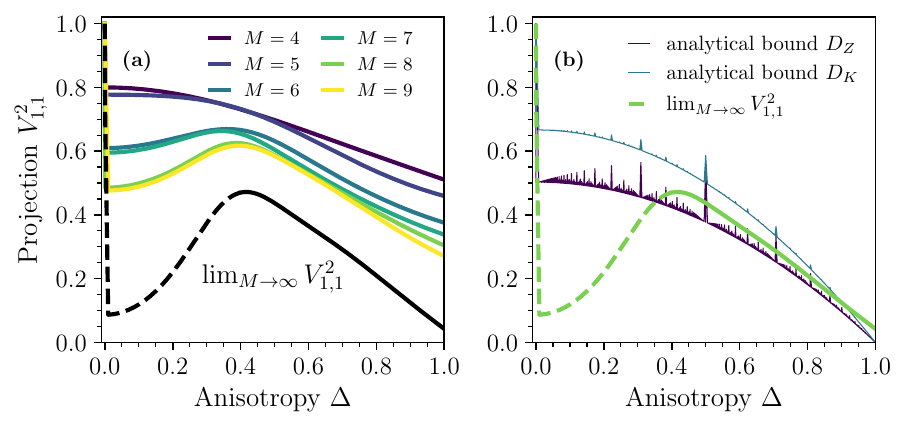}
    \caption{Projection of QLIOM, $\hat A^1$, on the spin current for the same case as in Fig.~\ref{fig:quasilocal}. \textbf{(a)} shows results for various $M$ and extrapolated results for $M\to \infty$. \textbf{(b)} compares extrapolated results with  Mazur bounds for spin current, $D_Z$ and $D_K$, obtained analytically in Refs.~\cite{Prosen2011a} and  ~\cite{Prosen2013a}, respectively. }
    \label{fig:stiff_bound2}
\end{figure}

Next, we focus on the extrapolated value $\lim_{M\to \infty} V^2_{1,1}$ which, according to \eqref{mazur}, is the contribution of the QLIOM, $\hat A^1$, to the spin stiffness. We have estimated the latter limit by fitting $V^2_{1,1}$ with a function linear in $1/M$, as shown in Fig.~\ref{fig:quasilocal}(b), and repeated similar calculations for other $\Delta$. Figure~\ref{fig:stiff_bound2}(a) shows  $V^2_{1,1}$ as a function of $\Delta$ for several supports, together with the extrapolated value for $M\to \infty$. Finally, in Fig.~\ref{fig:stiff_bound2}(b), we compare the estimated contribution to the spin stiffness with analytical predictions from Refs.~\cite{Prosen2011a} and \cite{Prosen2013a} which account for a single quasilocal conserved quantity ($D_Z$) and an uncountable family of quasilocal conserved quantities ($D_K$), respectively. One observes that $\hat A^1$ accurately reproduces the spin stiffness in the XXZ model down to $\Delta \simeq 0.4$. Note also that this particular QLIOM is insufficient to account for the stiffness for smaller values of $\Delta$. Figures~\ref{fig:quasilocal}(c) and (d) show that other QLIOMs also become important in this regime. However, in the case of these QLIOMs, the dependence of $\lambda_{\alpha}$ and $V^2_{\alpha,1}$ on $M$ for small $\Delta$ is not as smooth as for $\Delta >0.4$. Therefore, we do not formulate any quantitative conclusions about the $M\to \infty$ limit. We stress that our bound on the spin stiffness in Fig.  \ref{fig:stiff_bound2}(b) has been obtained for an infinite lattice (without diagonalizing any Hamiltonian) from a general method that is model-independent.

\subsection{Integrable unitary circuit} 
As a last example, we consider the simplest integrable unitary quantum circuit, obtained from the integrable trotterization of the XXX Heisenberg chain~\cite{Vanicat2018}. It is a brickwall quantum circuit, constructed from a two-site unitary matrix 
\begin{equation}
   \hat U_{l,l+1}(\delta) = \frac{ 1 + i \frac{\delta}{2} (1 + \hat{h}_{l,l+1})}{1+i\delta} \,,
    \label{eq:gate}
\end{equation}
 with \( \hat{h}_{l,l+1} =  1/2\left[ \hat{\sigma}^+_l \hat{\sigma}^-_{l+1} + \hat{\sigma}^-_l\hat{\sigma}^+_{l+1} \right] + \hat{\sigma}^z_l\hat{\sigma}^z_{l+1}\) being the local density of the XXX Hamiltonian. The  one-step unitary propagator is then defined as 
\begin{eqnarray}
   \hat{\mathcal{U}}(\delta) = \prod_{l=1}^{L/2} \hat{U}_{2l-1,2l}(\delta) \;\; \prod_{l=1}^{L/2}\hat{U}_{2l,2l+1}(\delta)
\end{eqnarray}
It consists of two half-steps, acting first on all even-odd, and next on all odd-even nearest neighbors. Although \(\hat{\mathcal{U}}(\delta)\) alone is manifestly nonlocal, \(\hat{\mathcal{U}}(\delta)^\dagger \hat O^s \hat{\mathcal{U}}(\delta) \) is still a local operator.

\begin{figure}
    \centering
    \includegraphics[width=1.0\linewidth]{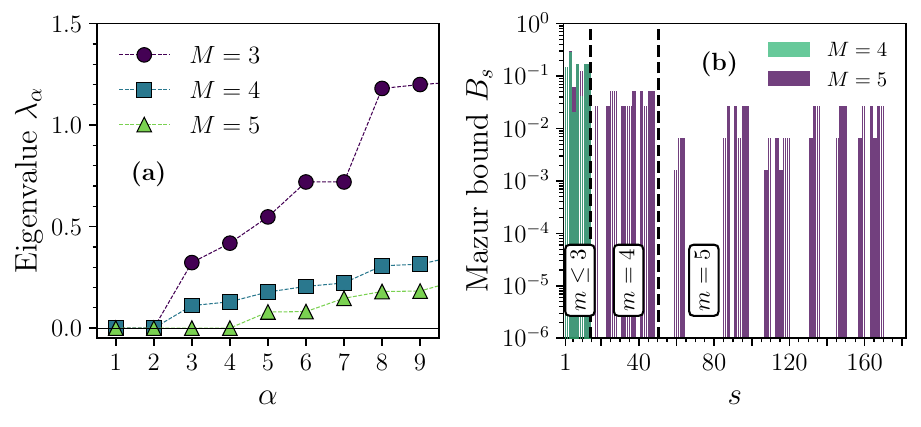}
    \caption{Solution of \eqref{eig} for the XXX quantum circuit defined via elementary gate in \eqref{eq:gate} with $\delta=0.5$.  \textbf{(a)} shows the eigenvalues and \textbf{(b)} shows the structure of LIOMs visualized via the Mazur bound from \eqref{mazur},  $B_s=\sum_{\alpha=1}^{N_L} V^2_{\alpha s}$. The index $s$ enumerates local
operators, $\hat O_s$, sorted accordingly to their support $m$.}
    \label{fig:circuit}
\end{figure}

For every LIOM \(\hat Q_n\) from continuous dynamics, we obtain two LIOMs \(\hat Q_n^{\pm}(\delta)\) in discrete dynamics, related to each other by single-site translation~\cite{Vernier2023}. Both \(\hat Q_n^{\pm}(\delta)\) are supported on \(2n+1\) sites, and in the limit \(\delta\to0\) reproduce XXX LIOMs. This observation allows us to choose an appropriate operator basis for our algorithm. We start with the same local basis as in the XXZ chain case, but consider extensive operators invariant under two-site translations, \(\hat{O}^{ {s},-} = \sum_l \hat{O}^{ {s}}_{2l-1}\) and \(\hat{O}^{ {s},+} = \sum_l \hat{O}^{ {s}}_{2l}\).
Due to the staggered structure of the propagator \(\hat{\mathcal{U}}(\delta)\), we have to relax the assumption that the first operator in the support is different from \(\hat{\sigma}^0_l\). We avoid trivial LIOMs related to magnetization conservation by removing \(m=1\) operators from the basis. 

Fig.~\ref{fig:circuit}(a) shows  eigenvalues obtained from \eqref{eig} while Fig.~\ref{fig:circuit}(b) demonstrates the structure of LIOMs encoded in the Mazur bounds $B_s$.   We observe two LIOMs for $M=3$. Extending the maximal support from
$M=3$ to $M=4$ neither yields additional LIOMs nor alters (up to numerical precision $\sim 10^{-12}$) the structure of LIOMs obtained for $M=3$. The next two LIOMs emerge for $M=5$ in agreement with  Ref.~\cite{Vanicat2018}.

\subsection{Nearly integrable spin ladders} 
In this section, we show that our approach can also be used for studying more realistic systems, i.e., perturbed integrable models, where it identifies slowly relaxing modes.  One may expect that operators, $\hat A^{\alpha}$, with the longest relaxation times correspond to the smallest norms \mbox{$\lVert i[\hat H,\hat A^{\alpha}] \lVert^2$}, see also Ref.~\cite{Song2025}.  Therefore, eigenvectors obtained from \eqref{eig1} for the smallest $\lambda_{\alpha} >0 $ correspond to the slowest modes of nearly integrable models. Without having access to the eigenvalues of the Hamiltonian,  one cannot get accurate results for the  time-dependence. However, we demonstrate that $\lambda_{\alpha}$ is a reasonable estimate of the relaxation rate of $\hat Q^{\alpha}$.  To this end, we consider  $\hat Q^{\alpha}$ that is a LIOM obtained for the integrable Hamiltonian $\hat H$. Upon adding a perturbation, $g \hat H'$,  one may estimate that  $\lambda_{\alpha} \simeq \lVert i[\hat H+g \hat H',\hat Q^{\alpha}] \lVert^2= g^2 \lVert i[ \hat H',\hat Q^{\alpha}]\lVert^2 \propto g^2$. Such quadratic dependence on the perturbation strength is generally expected for the relaxation rates, $\Gamma^{\alpha}$. In order to show that $\Gamma_{\alpha}\simeq \lambda_{\alpha}$, we  study a two-leg spin ladder
\begin{align}
 \hat H_{\text{ladder}} &= \sum_{y=1}^2 \hat H_{y} + U \sum_{l = 1}^{L} \hat \sigma^{z}_{l,y=1} \hat \sigma^z_{l,y=2}\; , \label{ham} \\
 \hat{H}_{y} &= \sum_{l= 1}^{L} \frac{1}{2} \left[ \hat \sigma^{+}_{l,y} \hat \sigma^{-}_{l+1,y} + \hat \sigma^{-}_{l,y} \hat \sigma^{+}_{l+1,y}\right]\label{haml}
 + \Delta \hat \sigma^{z}_{l,y} \hat \sigma^{z}_{l,y}\;,  \nonumber 
\end{align}
where $\hat \sigma^{s}_{l,y}$ is the spin operator at site $l$ in leg $y=1,2$.  This model is integrable for $U=0$ when it represents two decoupled XXZ chains, but also for $\Delta=0$ when it represents the {Fermi-Hubbard model}. We have chosen this Hamiltonian because various relaxation times in the nearly integrable regime ($0<\Delta \ll 1$ and $0<U \ll 1$) may differ by a few orders of magnitude and some of them are extremely long~\cite{Pawlowski2024}. For this reason,  numerical studies of the time evolution are very challenging.  

\begin{figure}[htbp]
  \centering
  \includegraphics[width=1\linewidth]{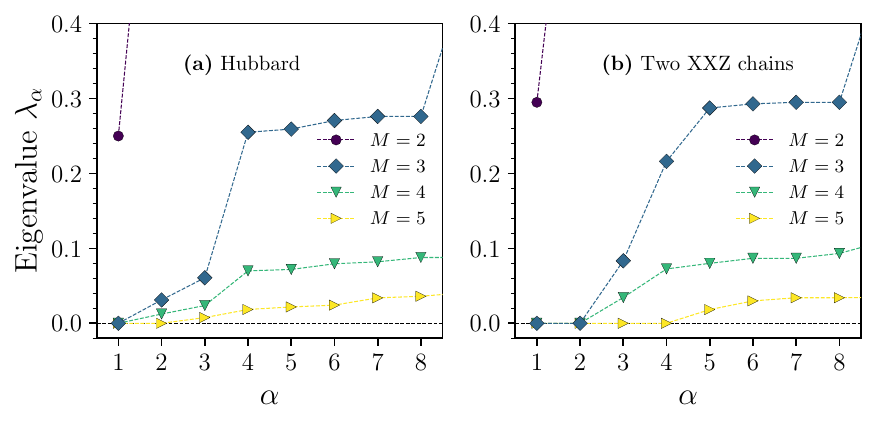}
  \caption{Smallest eigenvalues obtained from \eqref{eig},
for \textbf{(a)} Hubbard model, [\eqref{ham} with $\Delta=0$, $U=0.3$], and \textbf{(b)} two uncoupled XXZ chains [\eqref{ham} with $\Delta=0.3$ and $U=0$]. Operator basis was restricted to the symmetry sector supporting energy currents of uncoupled XXZ chains.}
 \label{fig:hubb}
\end{figure}  

For each leg, we use operators that are even under spin-flip and conserve the total $\hat{S}^z$. Then, we consider all possible products of densities that are odd under time-reversal, \(\hat O^{{s}}_{l,y=1,R} \hat O^{{s'}}_{l,y=2,I}\) and \(\hat O^{{s}}_{l,y=1,I} \hat O^{{s'}}_{l,y=2,R}\), and build the basis of extensive operators, $\hat O^{ss'}$. This symmetry sector includes energy currents of the uncoupled XXZ chains, denoted as $\hat Q^3_{y=1}$ and $\hat Q^3_{y=2}$. Before studying the perturbed model, let us discuss the integrable case.  Figure~\ref{fig:hubb} shows the smallest eigenvalues for $M\le 5$ (per each leg) obtained from \eqref{eig}  for $\Delta=0$ (left panel) and $U=0$ (right panel). For a fixed support, $M$, one observes that the decoupled XXZ chains have twice the number of LIOMs compared to the Hubbard model. 

\begin{figure}
    \centering
    \includegraphics[width=1.0\linewidth]{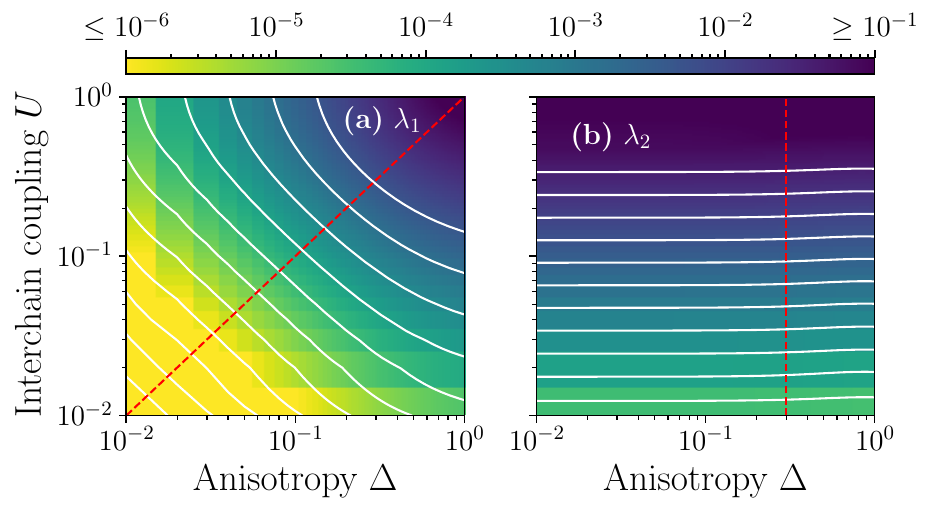}
    \caption{ Heatmaps of the smallest eigenvalue obtained from Eq.(\ref{eig}) for spin ladder defined in Eq.(\ref{ham}). Continuous white curves represent isolines.}
    \label{fig:ladder}
\end{figure}

Fig.~\ref{fig:ladder}  shows two smallest eigenvalues, $\lambda_1$ and $\lambda_2$, obtained from \eqref{eig} for the support $M=3$ and various values of $U$ and $\Delta$.  The shape of isolines (white curves) demonstrates that these eigenvalues exhibit very different dependence on the perturbation-breaking parameters. We observe that $\lambda_1$ is determined by the product $\Delta U$, and vanishes on both integrable lines. Therefore, the corresponding operator, $\hat A^1$, smoothly interpolates between the Hubbard LIOM, $\hat Q^3_H$, and the LIOM of the XXZ chains, $\hat Q^{+}=Q^3_{y=1}+Q^3_{2=1}$. Such smooth interpolation is possible because $\hat Q^{+}$ is very similar to $\hat Q^3_H$ provided that $U$ and $\Delta$ are small. In contrast to that, the second eigenvalue, $\lambda_2$, is determined by $U$ and vanishes only for $U=0$. Therefore, the corresponding operator, $\hat A^2$, represents a LIOM of uncoupled XXZ chains. However, this LIOM has no counterpart in the Hubbard model. We have found that \mbox{$\hat A^2\simeq \hat Q^{-} = Q^3_{y=1}-Q^3_{2=1}$}. We note that the different behavior of $\lambda_1$ and $\lambda_2$ shown in Fig.~{\ref{fig:ladder}} complies with the observation that the decoupled XXZ chains host twice as many LIOMs as the Hubbard chain. All Hubbard LIOMs can be deformed to the LIOMs of the XXZ chains. However, half of the LIOMs in the XXZ chains have no counterparts in the Hubbard model. 
 
\begin{figure}[htbp]
    \centering
    \includegraphics[width=1\linewidth]{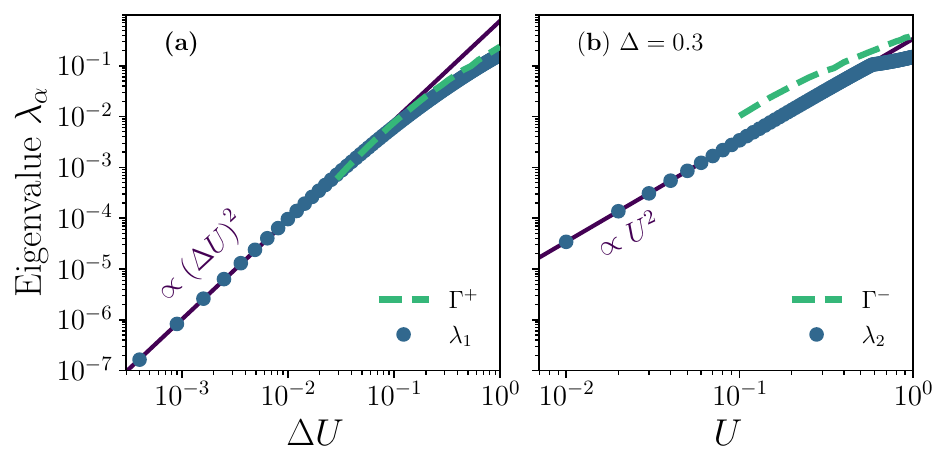}
    \caption{Cross-section of results from Fig.~\ref{fig:ladder} along the red lines shown in Fig.~\ref{fig:ladder}. $\Gamma^{\pm}$ shows the relaxation rate of $\hat Q^{\pm}=Q^3_{y=1} \pm Q^3_{2=1}$ estimated from Lanczos propagation method in Ref.~\cite{Pawlowski2024}. }
    \label{fig:relaxation}
\end{figure}

Figure~\ref{fig:relaxation} shows $\lambda_1$ and $\lambda_2$ calculated along the characteristic lines marked in Fig.~\ref{fig:ladder}. These results are compared with relaxation rates $\Gamma^+$ and $\Gamma^-$ roughly estimated  in Ref.~\cite{Pawlowski2024} from the decay of the correlation functions $\langle Q^{\pm}(t)  Q^{\pm} \rangle $. These correlation functions were obtained from the Lanczos propagation method for a ladder with $L=14$ rungs. The finite-size scaling was not possible, and weaker perturbations were numerically inaccessible. It is quite evident that the present approach overcomes limitations that occur in numerical studies based on time propagation.  We note that $\lambda_1 \simeq \Gamma^+ $ and $\lambda_2 \simeq \Gamma^-$, however, some differences are visible.   To explain the differences between $\lambda_{(1,2)}$ and $\Gamma^{\pm}$, we note that the present method selects the actual slow modes. The values of $\Gamma^{\pm}$ were calculated for fixed and guessed observables, $\hat Q^{\pm}$, which differ from the true slowest modes. Consequently, it is not surprising that $\lambda_1 < \Gamma^+$ and $\lambda_2 < \Gamma^-$.

\subsection{Slow modes in perturbed XXZ chain}
Upon adding a perturbation to the integrable model, all but a few LIOMs are destroyed, and they acquire finite relaxation times. However, one expects that the LIOMs of the parent integrable model remain the slowest modes, provided that the perturbation is sufficiently weak. In this sense, a weakly perturbed system retains information about the parent integrable model \cite{Kollar2011,bertini2015,mallayya2018}. A crucial question is whether one can identify the threshold perturbation strength at which this picture breaks down and determine the slowest mode that emerges for stronger perturbations. In this section, we identify such modes for perturbed easy-plane XXZ model, \mbox{$\hat H=\hat H_{\mathrm{XXZ}}+ \Delta_2 \hat H' $} with the perturbation $\hat H'=\sum_l \hat \sigma_l^z \hat  \sigma_{l+2}^z $. 

\begin{figure}[htbp]
    \centering
    \includegraphics[width=1\linewidth]{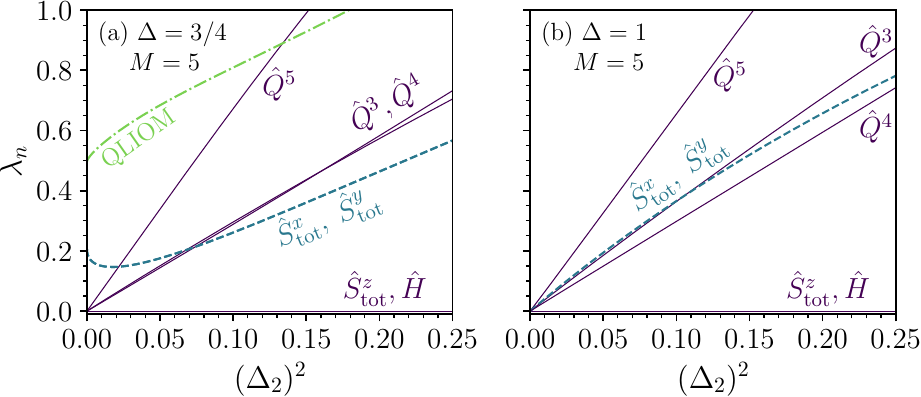}
    \caption{ Smallest eigenvalues obtained from \eqref{eig} for perturbed XXZ models at (a)
    $\Delta=3/4$ and (b) $\Delta=1$. Attached labels link the slow modes to the LIOMs of the parent integrable XXZ chain and QLIOM corresponds to the smallest $\lambda_{\alpha}$ in Fig.~\ref{fig:quasilocal}(a).
    }
    \label{fig:slow}
\end{figure}

Figure \ref{fig:slow}(a) shows the smallest eigenvalues obtained from \eqref{eig} for $\Delta=3/4$ as a function of the perturbation strength, $\Delta_2$. To avoid imposing any restrictions on the slow modes, the calculations were carried out in the Pauli string basis, and we included all operators supported on up to $M=5$ sites. Obviously, the perturbed model preserves two LIOMs corresponding to $\hat S^z_{\rm tot}$ and $\hat H$. All eigenvalues, $\lambda_{\alpha}$, that originate from other LIOMS of the parent model, increase smoothly with $\Delta_2$. Eventually, for $\Delta_2 >\Delta^{*}_2$, they exceed the relaxation rate of a mode that was slow but not conserved in the parent model.  This slowest mode for $\Delta_2 >\Delta^{*}_2$ corresponds to $\hat S^x_{\rm tot}$ and $\hat S^y_{\rm tot}$. We stress that the parent XXZ model at $\Delta=3/4$ as well as the perturbation, $\hat H'$, break the SU(2) symmetry. Still, the slowest mode is a LIOM of a nearby SU(2) symmetric XXX model, for which the results are shown in Fig.~\ref{fig:slow}(b). The closer $\Delta$ is to one, the weaker the perturbation at which $\hat S^{(x,y)}_{\rm tot}$ become relevant. Quite surprisingly, the subsequent slow mode is a QLIOM, which contributes to the spin stiffness for $\Delta< 1$ but does not exist for $\Delta\ge 1$. 

A nontrivial conclusion emerges from the above consideration: for a generic nonintegrable system, two slow modes cannot be linked (in principle) to a single integrable point. Typically, one attempts to relate the perturbed model to a single integrable parent model, and such a connection may be established, for example, via adiabatic flows \cite{Hyeongjin2024}. However, upon adding a perturbation, one does not depart from a single integrable point but rather from an integrable line in the parameter space. The latter line contains high-symmetry points (e.g., $\Delta=1$) that host more LIOMs (e.g., $\hat S^x_{\rm tot}$ and $\hat S^y_{\rm tot}$), which may be more robust against the perturbation. Consequently, the slow modes in perturbed systems are not inherited just from a single integrable Hamiltonian, but rather from various LIOMs that exist on the integrable line, $\Delta_2=0$.

\section{Concluding remarks} We have derived and implemented a general numerical approach that provides exact translationally-invariant  LIOMs and the corresponding Mazur bounds for infinite quantum systems for which the dynamics is set by either local Hamiltonian or by local unitary transformations. We can obtain the exact LIOMs for the thermodynamic limit because our approach, similarly to Refs.~\cite{Huse2015,Bulchandani2022,Song2025}, is targeting the short-time dynamics (as encoded, e.g., in the commutators $[\hat O, \hat H]$) under which local operator $\hat O$ remains local. Consequently, our results have an important advantage over approaches focused on the infinite-time properties, which are encoded in the diagonal matrix elements of observables calculated in the basis of the Hamiltonian's eigenstates. In the latter case, the thermodynamic limit is unreachable, and finite-size effects seem unavoidable. 

Diagonalization of the Hamiltonian (exact or approximate) is a common bottleneck for most problems studied in the context of interacting quantum systems. However,  our approach is not based on the diagonalization of the Hamiltonian or a unitary matrix. Therefore, it can be applied to problems that are hardly accessible to other numerical methods. In particular, it enables the study of problems beyond one-dimensional chains. It also identifies quasilocal conserved quantities in the XXZ chain and LIOMs in much more demanding system, i.e., in the Hubbard model. Furthermore, we have demonstrated that this approach also applies to nearly integrable systems, where it singles out slowly decaying observables and provides reasonable estimates of the corresponding relaxation rates. Studying the perturbed XXZ chain, we have found that these slow modes are not inherited from a single integrable Hamiltonian but rather from LIOMs which exist in a broader class of integrable models.  

It is rather obvious that one may also carry out similar calculations for systems that are not translationally invariant, e.g., containing inhomogeneous potentials. In the latter case, although one cannot study infinite systems anymore, the accessible system sizes are still much larger than those in approaches based on the diagonalization procedure.

Like all numerical methods, our approach also has limitations. This bottleneck concerns the number of local observables, $N_O\sim 10^4-10^5$, from which LIOMs are built by solving the eigenproblem in \eqref{eig}. In the case of spin systems, we still considered all operators supported on up to 10 consecutive sites. On the one hand, one may expect that finding very complex LIOMs may be problematic. On the other hand, such LIOMs only have a limited influence on the dynamics of physically relevant observables, which are typically represented by few-body local operators.

\vskip 1truecm  
\nocite{zenodo}
\acknowledgements{We acknowledge fruitful discussions with Marko  {\v Z}nidari{\v c} and Vir B. Bulchandani. J.P. acknowledges support by the National Science Centre (NCN), Poland via project 2023/49/N/ST3/01033. M.M. acknowledges support by the National Science Centre (NCN), Poland via project 2020/37/B/ST3/00020. 
}

\bibliography{manuscript}
\end{document}